# A Model of Entropy Production

Andreas Schlatter[1] (schlatter.a@bluewin.ch) and *R. E. Kastner[1,2](rkastner@umd.edu)

(1)The Quantum Institute, NY; (2) University of Maryland, College Park

11/1/24

*Abstract. A key tenet of the Transactional Interpretation of Quantum Mechanics is the idea that photon absorption localizes the absorbing material system. In doing so, it measures the location of the absorber and hence reduces information entropy which in turn needs to be balanced by appropriate entropy production, if there is a link between information entropy and thermodynamic entropy. Based on a critical analysis of the physics of information erasure, we clarify the link between information and thermodynamic entropy and develop a rigorous model of entropy production in photon-absorption processes. Links to the interpretation of quantum probabilities and to Maxwell's demon are made.*

1. Introduction

While the concept of entropy was introduced in thermodynamics by diverse scientists in the 19th century, the notion of information entropy is much more recent and goes back to a large extent to the seminal work of Claude Shannon. The question of how the two types of entropy are related has been discussed at length since the attempts to exorcise Maxwell's demon by Szilard [1] and Brillouin [2]. In their work it was by the process of measurement, i.e., the acquisition of knowledge, that the necessary entropy was produced to save the second law. Later, in connection with irreversible logical operations in computers, the thinking shifted to the view of Landauer [3], Bennett [4] i.e., that it is the erasure of information, stored in a memory, that is responsible for the necessary entropy production. This view runs in the literature under the name of Landauer's Principle and is pretty much the Received View today. Information entropy and thermodynamic entropy are thus often just identified, by transforming one into the other by means of multiplication by Boltzmann's constant. There is a smaller community of critiques, however; see e.g. [5], [6], [7].

There is another issue, often neglected in the literature: namely the fact that thermodynamic entropy is defined in connection with ensembles, whereas in quantum mechanics there is information entropy attributed to a single system in form of the Shannon entropy of its wave function. A measurement with a unique outcome therefore produces a reduction of information entropy in a single-trial, which must be compensated, if there is an identity of the two types of entropy. Just remaining at the level of expectation values and glossing over the single-trial falls short of saving the second law, which holds by its very definition for individual processes. Once taking the single trial into account, there arises the question whether the quantum probabilities are of epistemic or ontic nature, i.e. whether the probabilities merely represent the ignorance of the experimenter or whether they describe instrinsic uncertainty of the properties of physical systems. The Received View tacitly takes the classical, epistemic road, when for instance discussing the standard example of a one-bit system in form of a particle in a box [8].

Assume there is a box of volume $V$ and a particle in an unknown state which can with equal probability of $p = \frac{1}{2}$, say, be either of two, namely $|L\rangle$, meaning that the particle is in the left half of the box, or $|R\rangle$, meaning that it is in the right half. Box and particle are in equilibrium with a heat bath of temperature $T$. To reset the device, we introduce a frictionless piston from the right which moves until the middle of the box, $\frac{V}{2}$, and hence restores an initial situation where the particle is in state $|L\rangle$ with certainty. Now, for the energy difference of the total system there holds by the first law $dE = dW + dQ$. Since the process is supposed to be isothermal, we have no change of internal energy, $dQ = 0$, and $dE = dW = -pdV$, where $p$ is the pressure, which the piston exercises against the particle. Pressure can be written as an energy density and, since we are in thermal equilibrium, there is a particle energy of $k_B T$. There follows that $dE = -\frac{k_B T}{V} dV$ and that integrating from $V$ to $\frac{V}{2}$ leads to $E = k_B T \ln 2$, which implies an entropy production of $S = k_B \ln 2$. At the same time the information entropy of the system happens to be $I = -2\left(\frac{1}{2} \ln \frac{1}{2}\right) = \ln 2$ and hence $S = k_B I$. So, is Landauer's principle confirmed?

Let us look at the example once more. If the particle has in fact a definitive position, what does it matter whether we happen to know which state it is in, or not? If it is in fact in state $|L\rangle$, then the piston moves through empty space, pressure is zero and consequently $S = 0$. If it is in state $|R\rangle$, then, although there is work done by the piston to overcome the particle's inertia, there is no thermodynamic work $p\Delta V$, since the particle remains confined to a volume of $\frac{V}{2}$ and hence $dV = 0$ and equally $S = 0$. Likewise, if we just don't know where the particle is and want to measure it, then the same procedure leads to $S = 0$ if it is in state $|L\rangle$ and, if there is a sensor which stops the piston as soon as $p \neq 0$, then we measure the outcome $|R\rangle$ with arbitrarily small entropy production. This shows that, while Shannon entropy is reduced by the amount $ln2$ in any case, the physical state space of the particle is not; thus the epistemic view of quantum probabilities does not necessarily lead to thermodynamic entropy production, neither in case of resetting nor of measurement. This leaves us in a situation where there can be no general relation between information entropy and thermodynamic entropy. But is this the last word?

It is not, and in [7] it is argued that position measurements indeed produce thermodynamic entropy compensating the loss of information entropy in single-trials but only with a clear understanding that quantum probabilities are of ontic nature, i.e., that the particle really has no definite state before being measured. In this paper, we describe a specific measurement-process, namely the measurement of the position of a bound state by absorption of a photon, and give a detailed mathematical model for the argument in [7]. The model makes use of a specific form of position-momentum inequality.

2. The Model

Let there be a system $\Sigma_0$ consisting of a bound sate $\mathcal{B}$ in equilibrium with an environment of temperature $T_0$ and a photon $\gamma$ with energy $E_\gamma = h\nu$ before absorption by the bound state. To model the situation as simply as possible and reasonable, we assume that the wave function $\Psi(x, \vec{x}_j), j\epsilon J \subset \mathbb{N}$, of the bound state is factorizable as the product of a center of mass component $\psi(x)$ and an orbital component $Y(\vec{x}_j), j\epsilon J \subset \mathbb{N}, \Psi(x, \vec{x}_j) = \psi(x)Y(\vec{x}_j)$ [9]. Since the main contribution of the mass $m > 0$ stems from the nucleus, we may assume that the center of mass component $\psi(x)$ "carries" linear kinetic energy, whereas orbital energy components reside in $Y(\vec{x}_j)$. The bound state together with the photon form a closed system $\Sigma_0$ and under the assumption that the bound state "moves freely" with small momentum-uncertainty around a value $p_0$, the center of mass wave function can be assumed to be a Gaussian, $\psi_{p_0}(x)$ (i.e. $\left|\psi_{p_0}(x)\right|^2 \sim \mathcal{N}(\mu_x, \sigma_x)$). We choose as reference frame the direction of motion of the center of mass component in order to make calculations in one dimension possible. We will use $\psi_{p_0}(x)$ in order to model and analyze the entropic situation before and after absorption.

Let us define for any function $\psi \epsilon L^2(]-\infty, \infty[, \mathbb{C})$ the information entropy $I_\psi$ to be:

$$I_\psi = -\int_{-\infty}^{\infty} |\psi(s)|^2 \, ln|\psi(s)|^2 ds. \tag{1}$$

Given the center of mass wave function $\psi_{p_0}(x)$, the conjugate state $\varphi_{p_0}(p)$ is defined to be the Fourier transform $\varphi_{p_0}(p) = \hat{\psi}_{p_0}(p) = \frac{1}{\sqrt{2\pi\hbar}}\int_{-\infty}^{\infty} \psi_{p_0}(x)e^{-2\pi i p x}dx$. The total information-entropy $I_{\psi_{p_0}}^{tot}$ of $\psi_{p_0}(x)$ is then defined by the sum of the entropies of the center of mass component and its conjugate state:

$$I_{\psi_{p_0}}^{tot} = I_{\psi_{p_0}(x)} + I_{\varphi_{p_0}(p)}. \tag{2}$$

By a result of Leipnik [10] there holds for any pair of conjugate variables $\psi(x)$ and $\varphi(p)$ and with Planck's constant $h$:

$$I_\psi^{tot} = I_{\psi(x)} + I_{\varphi(p)} \geq ln\left(\frac{he}{2}\right), \tag{3}$$

with equality in case of Gaussian functions, which we may assume to be a good representation of systems in an equilibrium situation, as mentioned above. Since the bound state is supposed to move freely at a definite momentum, $\varphi_{p_0}(p)$ is highly concentrated around a mean value $\mu_p = p_0$ and there is hence a negative entropy contribution $I_{\varphi_{p_0}(p)} < 0$. Note that the differential entropy $I_\mathcal{N}$ of a Gaussian $\mathcal{N}(\mu, \sigma)$ is $I_\mathcal{N} = ln(\sqrt{2\pi}\sigma) + \frac{1}{2}$ and hence $\lim_{\sigma \searrow 0} I_\mathcal{N} = -\infty$. By (2) there holds:

$$I_{\varphi_{p_0}(p)} = ln\left(\frac{he}{2}\right) - I_{\psi_{p_0}(x)}. \tag{4}$$

At the same time, the momentum of the photon $\gamma$ is known to be $p_\gamma = \frac{h\nu}{c}$ and its position is undefinable, since there is no rest-frame. So, we have consistent with $\mu(p_\gamma) = 1$:

$$I_\gamma^{tot} = 0. \tag{5}$$

For the total system-entropy $I_{\Sigma_0}^{tot}$ before absorption we therefore have:

$$I_{\Sigma_0}^{tot} = I_{\psi_{p_0}}^{tot}. \tag{6}$$

Let us finally define, in analogy to Boltzmann's *H*-function, the thermodynamic entropy of the system $\Sigma_0$ by:

$$S^{\Sigma_0} = k_B I_{\varphi_{p_0}(p)} = -k_B \int_{-\infty}^{\infty} |\varphi_{p_0}(p)|^2 \ln|\varphi_{p_0}(p)|^2 dp, \tag{7}$$

where $k_B$ denotes the Boltzmann constant.

If initially we have $\psi_{p_0}(x) = \psi_{p_0}(x, 0)$ and $\varphi_{p_0}(p) = \varphi_{p_0}(p, 0)$, respectively, then a free evolution leads after some time $t > 0$ to new states $\psi_{p_0}(x, t)$ and $\varphi_{p_0}(p, t)$, still conjugates of each other (note that $\psi_{p_0}(x, t)$ is no longer a function with real variance). The evolution is unitary and causes an increasing dispersion, $\sigma_x(t)$, of the density $|\psi_{p_0}(x, t)|$ around some evolving position-mean value $\mu_x(t)$, while the density $|\varphi_{p_0}(p, t)|$ remains equally concentrated around $p_0$ and $|\varphi_{p_0}(p, t)| = |\varphi_{p_0}(p, 0)|$. Therefore, there holds by definition (7) for $t \geq 0$:

$$S^{\Sigma(t)} = const. \tag{8}$$

In other words, the entropy of the unitarily evolving free bound state remains constant, as expected from a reversible process (time reversal $t \to -t$ demands $\psi \to \psi^*$).

Let us now look at the situation after the absorption. The absorption at some time $t_1 > 0$ does two things at once: it annihilates the photon and localizes the center of mass component and thus transforms system $\Sigma_0$ into a spatially localized system $\Sigma_1$. This leaves us with a state $\psi_{x_1}(x)$, which is a Gaussian well concentrated around some spatial mean value $\mu_x = x_1$. So, there is now a negative entropy contribution $I_{\psi_{x_1}(x)} < 0$ to total entropy (2). But because of (3) the entropy-contribution of the conjugate Gaussian $\varphi_{x_1}(p)$ must compensate and we have in analogy to (4):

$$I_{\varphi_{x_1}(p)} = \ln\left(\frac{he}{2}\right) - I_{\psi_{x_1}(x)}. \tag{9}$$

So, by (4), (8) and (9) the transition $\Sigma(t) \to \Sigma_1$ induces for $0 \leq t < t_1$ an entropy difference of:

$$\Delta_{\Sigma(t)}^{\Sigma_1} S = k_B \left( I_{\varphi_{x_1}(p)} - I_{\varphi_{p_0}(p,t)} \right) = k_B \left( I_{\varphi_{x_1}(p)} - I_{\varphi_{p_0}(p,0)} \right) = k_B \left( I_{\psi_{p_0}(x)} - I_{\psi_{x_1}(x)} \right) > 0. \tag{10}$$

After the measurement, the bound state $\Sigma_1$ will again develop freely $\Sigma_1 \to \Sigma_1(t)$ and by equation (8) the entropy $S^{\Sigma_1(t)}$ remains constant, while the position state disperses around a moving mean-position.

Equation (10) is a consequence of quantum mechanics and holds in particular for a single system. How was it possible to detect entropy increase in closed systems in the 19th century? Evidently for (9) to hold, $\varphi_{x_1}(p)$ can no longer be concentrated around a momentum mean value. Indeed, we know that next to (3) there is another, related inequality for conjugate states $\psi(x), \varphi(p)$, namely the Heisenberg uncertainty inequality for the standard deviations $\sigma_x^\psi, \sigma_p^\varphi$:

$$\sigma_x^\psi \cdot \sigma_p^\varphi \geq \frac{\hbar}{2}, \tag{11}$$

with equality again in case of Gaussian functions. Hence, if the position-variance becomes small by the localization, then the momentum-variance increases the same way, as with negative position-entropy there comes along higher momentum-entropy (9). In the statistics of a classical ensemble, it is this fact (11), which is responsible for the observation that with localization there goes non-decreasing phase-space volume (this is the definition of entropy given by the expression $S = -k_B \log \Omega$, where $\Omega$ denotes phase-space volume). Therefore, it is due to equations (10) and (11) that we can understand classical phase-space thermodynamics, although it is in fact ontic uncertainty that supports the key observation. Hence, we can indeed formulate the following fundamental facts regarding the relationship between information-entropy, $I$, and thermodynamic entropy $S$ in absorption processes:

$$1)\ S = k_B \cdot I_{\varphi(p)} = k_B \int_{-\infty}^{\infty} |\varphi(p)|^2 \ln|\varphi(p)|^2 dp \qquad (12)$$

$$2)\ \Delta S \geq 0.$$

Note that the increase in momentum-spread of the bound state finds its physical explanation in the absorption of the momentum $p_\gamma = \frac{E_\gamma}{c}$ of the photon $\gamma$, consistent with the well-known fact that, to reach high resolution high energy is needed. Brioullin [2] rightly notes that, in order to distinctly localize the system $\Sigma$, the energy $E_\gamma$ of the photon has to satisfy $h\nu \gg k_B T_0$.

### 3. Some consequences

The process of absorptions might appear very specific and hence of limited consequence. But in [11] it was recently shown how this type of process and the corresponding entropy production can in fact be considered at the basis of our empirical universe and are possibly the true source of gravity. So, it might indeed be the "quantum-jumps" in localizations that are responsible for thermodynamic entropy production in the universe and (12) is arguably behind the second law, as Dirac mused early on [12]. Thermodynamic entropy in the universe cannot decrease, regardless of its initial level, since the universe is a priori a closed system instantiated in its parts by photon-absorptions [11].

Let us finally consider Maxwell's demon. The demon is a classical being and its construction is based on processing classical particles. But for a quantum particle there holds equation (12) and this opens the road to the exorcism. By (11) it becomes clear that, if the momentum of a particle is known, then there can be no sorting by letting fast particles pass through an opening, since the particle's position is totally uncertain. So, there is no "temperature-demon" [13]. On the other hand, also the kind of "pressure-demon" [13], where a sorting happens just by randomly letting particles pass from one compartment to the other unidirectionally, does not reduce entropy, since the implicit localization in passing an opening produces the necessary entropy to compensate (10). So, our considerations lead back to an original approach to explain away Maxwell's demon where the demon must obey the laws of quantum physics and where measurement is properly considered to be the source of entropy production to save the second law [1], [2].

### 4. Conclusion

We have developed a rigorous model which defines the relationship between information entropy and thermodynamic entropy in position measurements by photon absorption and proves that a localization necessarily produces the thermodynamic entropy $\Delta S$ to compensate the reduction of information entropy. We have found that there holds:

$$\Delta S \geq k_B \Delta I_{\varphi(p)}, \qquad (13)$$

where $I_{\varphi(P)}$ denotes the $H$-function. The model is entirely of quantum nature and holds for a single trial and hence suggests an ontic interpretation of the quantum probabilities, because it is by no means intelligible why the state of our knowledge should cause nature to reduce phase-space in a single experiment [5], [6], [7]. It revives the original view that an exorcism of Maxwell's demon rests on what happens when information is acquired by measurement, rather than erased in a memory. Generally however, with regard to other physical quantities, like spin, we are not aware of any rigorous proof of the identity of the corresponding information-and thermodynamic entropies.

Acknowledgements. The authors would like to thank two anonymous reviewers for valuable comments.

No datasets were generated or analysed during the current study.